\documentstyle[amssymb,aps,12pt]{revtex}
\begin{document}
\draft
\author{Sergio De Filippo\thanks{%
Unit\`{a} I.N.F.M., I.N.F.N. Salerno}}
\address{defilippo@sa.infn.it}
\author{Filippo Maimone\footnotemark[1] }
\address{maimone@sa.infn.it}
\author{Anton Luca Robustelli}
\address{alrobustelli@yahoo.com}
\address{Dipartimento di Fisica ''E.R. Caianiello'', Universit\`{a} di Salerno\\
Via Allende I-84081 Baronissi (SA) ITALY}
\date{\today}
\title{Numerical simulation of nonunitary gravity-induced localization}
\maketitle

\begin{abstract}
The localization of a quantum state is numerically exhibited in a nonunitary
Newtonian model for gravity. It is shown that an unlocalized state of a ball
of mass just above the expected threshold of $10^{11}\;$proton masses
evolves into a mixed state with vanishing coherences above some localization
lengths.
\end{abstract}


\section{Introduction.}

The apparently inescapable conclusion of the so called ''information loss
paradox'' was the assumption that a full theory of quantum gravity would be
non-unitary, with pure states evolving in principle into mixed ones\cite
{hawking1}. At the basis of such assumption was the observation that the
area of the event horizon of black holes increases monotonically in time
while information is irretrievably lost when matter falls into black holes.
This in turn led Bekenstein to propose that a generalized second law holds,
where the entropy of a black hole is proportional to the area of its event
horizon\cite{bekenstein}. More generally the laws of black hole dynamics
exhibit a remarkable resemblance with the laws of thermodynamics, if energy
is replaced by the black hole mass and temperature is assumed to be
proportional to the surface gravity. This idea was dramatically confirmed by
Hawking's discovery that black hole evaporation has a thermal character with
a temperature proportional to the surface gravity\cite{hawking2}.

It was argued that a non-unitary evolution should be effective even in
ordinary laboratory physics, since a continuous creation and evaporation of
submicroscopic black holes is expected within such a theory, even though it
seemed to give rise to unacceptably large violations of causality or
energy-momentum conservation\cite{ellis,banks}. A more recent analysis
pointed out that, in general, violations could be undetectable at the scales
of laboratory physics, and in particular that there need not be any (easily
detectable) conflict between non-unitarity, causality and energy
conservation if the model is non-Markovian, as suggested by black hole
dynamics\cite{unruh}. This analysis, though inspired by black hole dynamics,
was performed for generic non-unitary models without any specific reference
to gravity and intended only to mimic a possible effective model for gravity
at low energy.

It is remarkable that indications for a non-unitary evolution induced by
gravity also emerge in the very context of non-relativistic QM, in which
many tentative ad hoc modifications of the Schroedinger equation
(''reduction models'') have been proposed to account for both wave-function
collapse and the emergence of classicality\cite{ghirardi,pearle,diosi}. In
fact in these models, in order to minimize energy pumping one is induced to
couple the phenomenological stochastic external field mainly to the centre
of mass, which strongly suggests its gravitational origin\cite{pearle1,ring}.

In view of a possible convergence of these two research lines (information
loss paradox and reduction models), in some recent papers one of the authors
proposed and analyzed a non-Markovian non-unitary model for Newtonian
gravity \cite{defilippo1,defilippo3}. It has been shown that this model can
be derived from fourth-order gravity in which, while classical instability
is cured at expense of unitarity, the nice one-loop ultraviolet behavior of
HD gravity is preserved\cite{defilippo5,defmaim}.

While for the non-Markovian models considered by Unruh and Wald the basic
idea is to have the given system interacting with a ''hidden system'' with
''no energy of its own and therefore... not... available as either a net
source or a sink of energy'' \cite{unruh}, in the present model energy
conservation is granted by the ''hidden system'' being a copy of the
physical system, coupled to it only by gravity, and constrained to be in its
same state and then to have its same energy. The unitary dynamics and the
states referred to the doubled operator algebra are what we call
respectively meta-dynamics and meta-states, while, by tracing out the hidden
degrees of freedom, we get the non-unitary dynamics of the physical states.
Pure physical states correspond then to meta-states without entanglement
between physical and hidden degrees of freedom.

In particular it has been shown that, while reproducing the classical
aspects of the Newtonian interaction, the model gives rise to a threshold,
which for ordinary condensed matter densities corresponds to $\sim 10^{11}$
proton masses, above which self-localized center of mass wave functions
exist, in analogy to the Schroedinger-Newton model\cite
{christian,penrose1,moroz,tod}. However, while the latter produces rather
unphysical stationary localized states without spreading, according to the
present model an initial localized pure state evolves in time into an
unlocalized ensemble of localized states. That is consistent with the
expectation that (self-)gravity may produce a growing entropy in a genuinely
isolated system \cite{defilippo2,defmaim}, as suggested by black hole
dynamics.

It was also shown that the meta-dynamics can be reformulated in terms of a
functional integral over two auxiliary scalar fields. If the resulting
expression is applied to a physical state, obtained by tracing out the
hidden degrees of freedom, one gets an expression which can be taken as an
independent definition of the model, free from any reference to hidden
degrees of freedom\cite{defilippo3,defmaim}. This expression was used both
to show that the $N\rightarrow \infty $ limit of the $N$-hidden replicas
generalization of the model reproduces the Schroedinger-Newton model and to
analyze wave function reduction\cite{defilippo4}. To be more specific, the
superposition of a large number of widely spaced localized states,
corresponding to the ground meta-state of the relative motion of a physical
body of mass $M$ and its hidden partner, was considered. It was shown that,
within a random phase approximation and omitting the slow spreading of the
localized states, it evolves into an ensemble of the mentioned localized
states in a typical time $\tau _{g}=\hbar G^{-1}M^{-5/3}\varrho ^{-1/3}\sim
10^{20}\left( M/m_{p}\right) ^{-5/3}%
%TCIMACRO{\unit{s}}%
%BeginExpansion
\mathop{\rm s}%
%EndExpansion
$ (for ordinary matter density $\varrho \sim 10^{24}m_{p}/%
%TCIMACRO{\unit{cm}}%
%BeginExpansion
\mathop{\rm cm}%
%EndExpansion
^{3}$), where $G$ and $m_{p}$ respectively denote the gravitational constant
and the proton mass.

The key hypothesis behind any model of emergent classicality is that for
some reason (the surrounding environment or a fundamental noisy source)
unlocalized macroscopic states get localized, i.e. that coherences for space
points farther than some localization lengths vanish \cite{paz}. In such a
way embarrassing quantum superpositions of distinct position states of
'macroscopic' bodies are avoided, thus recovering an essential element of
reality of our classical realm of predictability: the elementary fact that
bodies are observed to occupy quite definite positions in space. The aim of
the present paper is to check that feature, which makes the model a viable
localization model, without using very peculiar initial conditions or any
approximation, and independently of the heuristic functional formulation. In
order to make the problem numerically tractable we take a rotational
invariant initial state of an isolated ball of ordinary matter density and a
total mass just above the mass threshold, namely in the least favorable
conditions to exhibit dynamical localization. The ensuing phenomenology may
be also relevant in view of possible future experimental checks of
fundamental (not environment induced) decoherence. In fact in the last few
years much efforts have been done in order to get a quantitative control of
the environmental decoherence \cite{zeilinger} and to construct cat states
involving growing masses. In this respect an improvement in the
Bose-Einstein Condensation (BEC) technology may be useful to test
gravity-induced decoherence, considering the unprecedented scale of
controlled quantum coherence achieved there.

\section{The model.}

We give here a concise definition of the model. Let $H[\psi ^{\dagger },\psi
]$ be the non-relativistic Hamiltonian of a finite number of particle
species, like electrons, nuclei, ions, atoms and/or molecules including also
the halved gravitational interaction, where $\psi ^{\dagger },\psi $ denote
the whole set $\psi _{j}^{\dagger }(x),\psi _{j}(x)$ of
creation-annihilation operators, i.e. one couple per particle species and
spin component. $H[\psi ^{\dagger },\psi ]$\ includes the usual
electromagnetic interactions accounted for in atomic, molecular and
condensed-matter physics. To incorporate that part of gravitational
interactions responsible for non-unitarity, one has to introduce
complementary creation-annihilation operators $\widetilde{\psi }%
_{j}^{\dagger }(x),\widetilde{\psi }_{j}(x)$ and the overall
(meta-)Hamiltonian 
\begin{equation}
H_{G}=H[\psi ^{\dagger },\psi ]+H[\widetilde{\psi }^{\dagger },\widetilde{%
\psi }]-\frac{G}{2}\sum_{j,k}m_{j}m_{k}\int dxdy\frac{\psi _{j}^{\dagger
}(x)\psi _{j}(x)\widetilde{\psi }_{k}^{\dagger }(y)\widetilde{\psi }_{k}(y)}{%
|x-y|}  \label{meta-hamiltonian}
\end{equation}
acting on the product $F_{\psi }\otimes F_{\widetilde{\psi }}$ of the Fock
spaces of the $\psi $ and $\widetilde{\psi }$ operators, where $m_{i}$ is
the mass of the $i$-th particle species and $G$ is the gravitational
constant. The $\widetilde{\psi }$ operators obey the same statistics as the
corresponding operators $\psi $, while $[\psi ,\widetilde{\psi }]_{-}=[\psi ,%
\widetilde{\psi }^{\dagger }]_{-}=0$.

The meta-particle state space $S$ is the subspace of $F_{\psi }\otimes F_{%
\widetilde{\psi }}$ including the meta-states obtained from the vacuum $%
\left| \left| 0\right\rangle \right\rangle =\left| 0\right\rangle _{\psi
}\otimes \left| 0\right\rangle _{\widetilde{\psi }}$ by applying operators
built in terms of the products $\psi _{j}^{\dagger }(x)\widetilde{\psi }%
_{j}^{\dagger }(y)$ and symmetrical with respect to the interchange $\psi
^{\dagger }\leftrightarrow \widetilde{\psi }^{\dagger }$, which, then, have
the same number of $\psi $ (physical) and $\widetilde{\psi }$ (hidden)
meta-particles of each species. As for the observable algebra, since
constrained meta-states cannot distinguish between physical and hidden
operators, it is identified with the physical operator algebra. In view of
this, expectation values can be evaluated by preliminarily tracing out the $%
\widetilde{\psi }$ operators. In particular, for instance, the most general
meta-state corresponding to one particle states is represented by 
\begin{equation}
\left| \left| f\right\rangle \right\rangle =\int dx\int dyf(x,y)\psi
_{j}^{\dagger }(x)\widetilde{\psi }_{j}^{\dagger }(y)\left| 0\right\rangle
,\;\;f(x,y)=f(y,x).  \label{f}
\end{equation}
This is a consistent definition since $H_{G}$\ generates a group of
(unitary) endomorphisms of $S$.

\section{Dynamical localization: numerical results.}

Consider now a uniform matter ball of mass $M$ and radius $R$. Within the
model the Schroedinger equation for the meta-state wave function $\Xi
(X,Y,t) $ is given by 
\begin{equation}
i\hslash \frac{\partial \Xi }{\partial t}=\left[ -\frac{\hslash ^{2}}{2M}%
(\nabla _{X}^{2}+\nabla _{Y}^{2})+V(\left| X-Y\right| )\right] \Xi
\label{schroedinger}
\end{equation}
where $X$ and $Y$ respectively denote the position of the center of mass of
the physical body and of its hidden partner, while $V$ is the (halved)
gravitational mutual potential energy of the two interpenetrating
meta-bodies, which, as can be shown by an elementary calculation, is 
\begin{equation}
V(r)=\frac{1}{2}GM^{2}\left( \frac{\theta
(2R-r)(80R^{3}r^{2}-30R^{2}r^{3}+r^{5}-192R^{5})}{160R^{6}}-\frac{\theta
(r-2R)}{r}\right)  \label{potential}
\end{equation}
where $\theta $ denotes the Heaviside function. Observe that our final
result can be immediately reread as the solution for the whole set of
parameters obtained by the scaling: 
\[
t\rightarrow \lambda t,\;\;\;\;\;M\rightarrow \lambda
^{-1/5}M,\;\;\;\;\;\;X\rightarrow \lambda ^{3/5}X,\;\;\;\ \;R\rightarrow
\lambda ^{3/5}R\;\;\;\;\;\; 
\]
\begin{equation}
\Xi \left( X,Y;t\right) \rightarrow \Xi \left( \lambda ^{-3/5}X,\lambda
^{-3/5}Y;t\right)
\end{equation}
where $\lambda $ is a real positive dimensionless parameter. This is
consistent with the expression for $\tau _{g}$, the latter giving $\tau
_{g}\rightarrow \lambda \tau _{g}$. Besides, for consistency, we note that
the mass cannot cross the threshold, as the latter scales with the same
power law of the mass itself $M_{t}\rightarrow \lambda ^{-1/5}M_{t}$.

If we separate Eq. (\ref{schroedinger}) into the equation for the relative
motion and that for the center of mass, then for ordinary matter density $%
\varrho \sim 10^{24}m_{p}/%
%TCIMACRO{\unit{cm}}%
%BeginExpansion
\mathop{\rm cm}%
%EndExpansion
^{3}$ and $M$ above the threshold $\sim 10^{11}m_{p}$, the former admits
bound meta-states of width $\Lambda _{G}\sim (m_{p}/M)^{1/2}cm$,
corresponding to small oscillations around the minimum of the gravitational
potential. In particular an untangled localized meta-state, corresponding to
a physical pure state is: 
\begin{equation}
\Psi _{TOT}=\Psi _{0}(X)\Psi _{0}(Y)=\Psi _{0}(\left[ X+Y\right] /2)\Psi
_{0}(\left[ X-Y\right] /2),  \label{pure}
\end{equation}
where 
\[
\Psi _{0}(\left[ X-Y\right] /2)=\left( \Lambda _{G}^{2}\pi \right)
^{-3/4}\exp [-\left| X-Y\right| ^{2}/(2\Lambda _{G}^{2})];\;\ \ \;\Lambda
_{G}=(8\hslash ^{2}R^{3}/GM^{3})^{1/4} 
\]
is proportional to the ground meta-state of the relative motion in the
hypothesis $\Lambda _{G}\ll R$.

If $\Xi (X,Y)\equiv \psi (\left[ X+Y\right] /2)\phi (X-Y)$, the equation for 
$\phi $, due to the spherical symmetry, reduces to the radial equation for $%
\chi (r)\equiv \phi (\left| X-Y\right| )$. This equation has been solved
numerically by the algorithm obtained from the space discretization over $%
10^{4}$ points of the equation 
\begin{equation}
\left[ 1+\frac{1}{2}iHdt/\hslash \right] u(r,t+dt)=\left[ 1-\frac{1}{2}%
iHdt/\hslash \right] u(r,t)+o(dt^{2})
\end{equation}
in the interval $(-R/2,R/2)$, where $H$ denotes the radial Hamiltonian for $%
u(r,t)\equiv r\chi (r,t)$.$\;$Such a procedure assures the stability of the
state-vector norm during the time evolution and is second-order accurate\cite
{recipes}.

A uniform ball of mass $M=0.38\times 10^{12}m_{p}$ and radius $R=4.8\times
10^{-5}%
%TCIMACRO{\unit{cm}}%
%BeginExpansion
\mathop{\rm cm}%
%EndExpansion
$ was considered, with initial conditions like in Eq.(\ref{pure}), but for $%
\Lambda _{G}\sim 1.6\times 10^{-6}%
%TCIMACRO{\unit{cm}}%
%BeginExpansion
\mathop{\rm cm}%
%EndExpansion
$ replaced by $\Lambda =5.6\Lambda _{G}$. The solution of Eq. (\ref
{schroedinger}) is then obtained as the product of $\phi $ and the
analytical solution 
\begin{equation}
\psi (\left[ X+Y\right] /2,t)\propto \exp \left[ \frac{-\left| X+Y\right|
^{2}/4}{\Lambda ^{2}/2+i\hslash t/M}\right] 
\end{equation}
of the center of meta-mass equation. 

The physical state $\rho $ is evaluated by tracing out the hidden body 
\begin{equation}
\rho (X;X^{^{\prime }})=\int dY\Xi (X,Y)\Xi ^{*}(X^{\prime },Y).
\label{phstate}
\end{equation}

After an evolution time $t=10%
%TCIMACRO{\unit{s}}%
%BeginExpansion
\mathop{\rm s}%
%EndExpansion
$ ($\sim \tau _{g}$) the function $\widetilde{\varrho }\left(
X_{1},X_{1}^{\prime }\right) \equiv \rho (X_{1},0,0;X_{1}^{^{\prime }},0,0)$
can be represented as in Fig. (4). To compare with the free evolution, in
which the dynamics gives the usual spreading, we have also shown the
corresponding function in Fig.(3). 

The final result can even be fitted by the product of two Gaussian
functions: 
\begin{equation}
\widetilde{\varrho }\left( X_{1},X_{1}^{\prime }\right) =\exp
[-(X_{1}+X_{1}^{^{\prime }})^{2}/\Lambda _{+}^{2}]\exp
[-(X_{1}-X_{1}^{^{\prime }})^{2}/\Lambda _{-}^{2}]  \label{state}
\end{equation}
where $\Lambda _{+}=0.27R\sim 1.3\times 10^{-5}%
%TCIMACRO{\unit{cm}}%
%BeginExpansion
\mathop{\rm cm}%
%EndExpansion
$ and $\Lambda _{-}=1.8\times 10^{-2}R\sim 8.1\times 10^{-7}%
%TCIMACRO{\unit{cm}}%
%BeginExpansion
\mathop{\rm cm}%
%EndExpansion
$, while the free evolution, ignoring the gravitational (self-)interaction,
would give the same product structure with $\Lambda _{+}=\Lambda _{-}\sim
1.3\times 10^{-5}%
%TCIMACRO{\unit{cm}}%
%BeginExpansion
\mathop{\rm cm}%
%EndExpansion
\;$(see Figs.5 and 6). In spite of the fact that the fit has been performed
by simple Gaussian functions, with the height and the size as independent
parameters, the result turns out to be very accurate. 

It should be remarked that the exact rotational invariance of the initial
state, which rather artificially leads to a degenerate final state, does not
play any special role. In fact, due to the unitarity in the enlarged Hilbert
space, modifying slightly the initial state would result in a slight
modification of the final metastate, and ultimately of the physical state.
Nevertheless the choice of a pure initial state corresponds to a precise
measurement, which represents in our case a rather ideal situation. On the
other hand the simplicity of the state constraint renders even conceptually
more precise such a measurement with respect to the case of a system
interacting with a complex environment, where the state of the environment
after the measurement is in general a complex functional of the system
state, to a large extent uncontrollable. In any case the present calculation
has to be considered only as a first step in the analysis of an intrinsic
nonunitary model.

\section{Entropy estimation and final remarks.}

The model may also be relevant to the quantum foundations of thermodynamics,
without resorting to the subjective and vaguely defined notion of coarse
graining, by the identification of both ordinary entropy and black hole
entropy with von Neumann entropy, i.e. with the entanglement entropy with
hidden degrees of freedom \cite{defmaim}. In fact a theory that is
intrinsically nonunitary, i.e. within which entropy can vary with time even
in closed systems, allows in principle a much more satisfactory derivation
of the thermodynamical tendency to equilibrium, assuming with Boltzmann an
appropriate choice of the initial conditions.

To estimate the final entropy production, consider that $\delta
(X_{2})\delta (X_{3})\widetilde{\varrho }(X_{1},X_{1}^{\prime })\delta
(X_{2}^{\prime })\delta (X_{3}^{\prime })$ represents the state resulting
from a measurement on $\rho $, giving $X_{2}=X_{3}=$ $0$, by which one means
that the uncertainties are smaller than the typical scale of variation of $%
\rho $. The entropy of $\widetilde{\varrho }$ then gives a lower bound for
the entropy of $\rho $. To estimate the entropy $S\left[ \widetilde{\varrho }%
\right] $ of $\widetilde{\varrho }$, we evaluate its purity 
\begin{equation}
Tr\widetilde{\varrho }^{2}=\int dX_{1}dX_{1}^{\prime }\widetilde{\varrho }%
\left( X_{1},X_{1}^{\prime }\right) \widetilde{\varrho }\left( X_{1}^{\prime
},X_{1}\right) ,
\end{equation}
which, for $\widetilde{\varrho }$ replaced by its analytical fit (Eq. \ref
{state}), gives $Tr\widetilde{\varrho }^{2}=\Lambda _{-}/\Lambda _{+}\sim
6.\times 10^{-2}$. If, for simplicity, we consider the corresponding
ensemble as one of $N$ equiprobable states, then 
\begin{equation}
Tr\widetilde{\varrho }^{2}=\sum_{j=1}^{N}\frac{1}{N^{2}}=\frac{1}{N}%
\Rightarrow N\sim 17;\;S\left[ \widetilde{\varrho }\right] \sim K_{B}\log 17.
\end{equation}
If we approximate $\rho \,$as the direct product of three equivalent $%
\widetilde{\varrho }$, namely we omit the entanglement between the three
Cartesian coordinates, which is absent in the initial pure state and would
stay so if the potential in Eq. (\ref{potential}) were replaced by its
quadratic approximation, then the total entropy is $S\left[ \rho \right] =3S%
\left[ \widetilde{\varrho }\right] $, corresponding to $N^{3}$ equiprobable
states.

Independently of any approximation $\rho (X;X^{^{\prime }})$, as the kernel
of a compact positive semi-definite Hermitian operator of unit trace\cite
{vonNeumann}, can be diagonalized as 
\begin{equation}
\rho (X;X^{^{\prime }})=\sum_{j}p_{j}\psi _{j}(X)\psi _{j}^{\ast }(X^{\prime
});\;p_{j}\geq 0;\;\sum_{j}p_{j}=1;\;\left\langle \psi _{j}|\psi
_{k}\right\rangle =\delta _{jk},
\end{equation}
where the above approximate estimate makes us expect that $%
-\sum_{j}p_{j}\log p_{j}\sim 3\log 17$. This result has to be compared with
a value $\sim 3\log \left[ \Lambda _{+}/\Lambda _{G}\right] \sim 3\log 8$
corresponding to a naive counting where the orthogonal states $\psi _{j}$
above are assumed localized and approximately non overlapping. This small
discrepancy is not surprising, as these states are expected to include
contributions from several low lying bound meta-states, so that they are
orthogonal in spite of the overlapping of their probability densities, due
to their space oscillations.

The qualitative agreement between our two estimates of entropy, the one
corresponding to the computed density matrix and the other corresponding to
the approximation of the mixed state by means of an ensemble of equiprobable
localized states which occupy the volume roughly occupied by the density $%
\rho (X;X)$, makes it natural to assume that the states $\psi _{j}$
diagonalizing $\rho $ are localized. To be more precise, this would be
strictly true, without ambiguities, after breaking the exact rotational
invariance, which, as mentioned, is expected to introduce artificial
degeneracies in the density operator.

The present result strengthens our confidence in the most relevant
peculiarities of the localization phenomenology ensuing from the model,
which make it {\it in principle} distinguishable both from the other
proposed models\cite{ghirardi,pearle} and possibly from the competing action
of the environment-induced decoherence\cite{paz}. In particular the model
presents a sharp threshold, below which localization is practically absent,
and a localization time $\tau _{g}\propto M^{-5/3}$ rapidly decreasing, as
the mass is increased. Furthermore the threshold mass $M_{t}\propto \rho
^{1/10}$ is remarkably robust with respect to mass density variations.

Acknowledgments - Financial support from M.U.R.S.T., Italy and I.N.F.M.,
I.N.F.N. Salerno is acknowledged.

\bigskip Captions :

Fig. 1. Initial radial wave function $\chi \left( r^{\prime }\right) $, with 
$r^{\prime }\equiv $ $r/R$, for the internal state of the two meta-bodies.

Fig. 2. Radial wave function $\chi \left( r^{\prime }\right) $, with $%
r^{\prime }\equiv $ $r/R$, for the internal state of the two meta-bodies at
the final simulation time $t=10%
%TCIMACRO{\unit{s}}%
%BeginExpansion
\mathop{\rm s}%
%EndExpansion
$.

Fig. 3. The modulus of the reduced density matrix $\left| \rho \left(
X,X^{\prime }\right) \right| $ in the absence of gravity is shown in the
plane $X_{1},X_{1}^{\prime }$ with $X_{2}=X_{2}^{\prime
}=X_{3}=X_{3}^{\prime }=0$,$\;$at the evolution time $t=10$ $%
%TCIMACRO{\unit{s}}%
%BeginExpansion
\mathop{\rm s}%
%EndExpansion
$.

Fig. 4. Modulus of the reduced density matrix $\left| \rho \left(
X,X^{\prime }\right) \right| \;$with the inclusion of gravity in the plane $%
X_{1},X_{1}^{\prime }$ with $X_{2}=X_{2}^{\prime }=X_{3}=X_{3}^{\prime }=0\;$%
at the evolution time $t=10$ $%
%TCIMACRO{\unit{s}}%
%BeginExpansion
\mathop{\rm s}%
%EndExpansion
$.

Fig. 5. A longitudinal Gaussian profile corresponding to the function $%
\left| \widetilde{\rho }\left( X,X\right) \right| $ has been superimposed to
the array of points obtained from the numerical simulation. Lengths $X$ on
the x-axis are measured in units of the radius $R$.

Fig. 6. Gaussian transverse cross-sections corresponding to the function $%
\left| \widetilde{\rho }\left( X,-X+k\ast 0.08R\right) \right| ,$ $\
k=0,1,2,3,4$ have been superimposed to the arrays of points obtained from
the numerical simulation. The dashed line represents the longitudinal cross
section. Lengths $X$ on the x-axis are measured in units of $10^{-2}R$.

\end{document}